# Knot Architecture for Biocompatible Semiconducting Two-Dimensional Electronic Fibre Transistors

*Tian Carey*[1*], *Jack Maughan*[1], *Luke Doolan*[1], *Eoin Caffrey*[1], *James Garcia*[1], *Shixin Liu*[1], *Harneet Kaur*[1], *Cansu Ilhan*[1], *Shayan Seyedin*[2], *Jonathan N. Coleman*[1*]

[1]*School of Physics, CRANN & AMBER Research Centres, Trinity College Dublin, Dublin 2, Ireland*

[2]*School of Engineering, Newcastle University, UK*

[*careyti@tcd.ie](mailto:careyti@tcd.ie), [colemaj@tcd.ie](mailto:colemaj@tcd.ie)

*Correspondence and requests for materials should be addressed to TC.

**In recent years, the rising demand for close interaction with electronic devices has led to a surge in the popularity of wearable gadgets. While wearable gadgets have generally been rigid due to their utilisation of silicon-based technologies, flexible semiconducting fibre-based transistors will be needed for future wearables as active sensing components or within microprocessors to manage and analyse data. Two-dimensional (2D) semiconducting flakes are yet to be investigated in fibre transistors but could offer a route toward high-mobility, biocompatible and flexible fibre-based devices. Here we report the electrochemical exfoliation of semiconducting two-dimensional (2D) flakes of tungsten diselenide ($WSe_2$) and molybdenum disulfide ($MoS_2$). The high aspect ratio (>100) of the flakes achieves aligned and conformal flake-to-flake junctions on polyester fibres enabling transistors with mobilities $\mu$ ~ 1 cm$^2$ V$^{-1}$ s$^{-1}$ and a current on/off ratio, $I_{on}/I_{off}$ ~ 10$^2$ - 10$^4$. Furthermore, the cytotoxic effects of the $MoS_2$ and $WSe_2$ flakes with human keratinocyte cells are investigated and found to be biocompatible. As an additional step, we create a unique transistor 'knot' architecture by leveraging the fibre diameter to establish the length of the transistor channel, facilitating a route to scale down transistor channel dimensions (~ 100 μm) and utilise it to make $MoS_2$ fibre transistors with a human hair that achieves mobilities as high as $\mu$ ~ 15 cm$^2$ V$^{-1}$ s$^{-1}$.**

There is an increasing trend toward incorporating technology into the clothes we wear. In just the last thirty years, over two-thirds of the world population have begun to carry a smartphone in their pocket,[1] and there is a growing demand for wearable devices with similar functionality.[2] However, many wearable technologies are still in their infancy and are based on bulky rigid silicon components that were never initially intended for use with textile or fibre substrates.[2] Therefore, new protocols, materials and device architectures will be needed to manufacture next-generation wearables so that the electronic devices can meet the stringent requirements of textile and fibre substrates, such as breathability, biocompatibility, flexibility, comfort, washability and conformability of the components to the textile substrate.[3] Wearable technology will transition from glasses, bands and watches to seamlessly embedded technology such as conformable fibres, textiles, tattoos or implantable devices that will have features such as disease monitoring,[4] motion tracking,[5] thermal regulation,[6] lighting[7] and energy harvesting components (e.g. photovoltaics, triboelectric, piezoelectric and thermoelectric nanogenerators).[8-10] In many of these applications, transistors will be used as an active data collection component or as part of a microprocessor to analyse data.[4]

Textile transistors can be made as planar structures on textiles[11] or along single fibres or fibrils.[12] Manufacturing transistors on fibres can be more challenging, often requiring transistor components smaller than the fibre diameter (<500 µm).[3] However, fibre-based devices can benefit from their three-dimensional structure, which permits the possibility of novel device architectures,[13] compatibility with current textile manufacturing processes and integration into multifunctional coaxial systems capable of performing multiple tasks simultaneously within a single fibre (e.g. energy generation and sensing).[14] Fibre transistors have been demonstrated with organic conducting polymers (*e.g.* poly(3,4-ethylenedioxythiophene) polystyrene sulfonate, PEDOT:PSS and Poly(3-hexylthiophene), P3HT), [12, 15-17] semiconducting carbon nanotubes (SCNT),[18] and metal oxides (*e.g.* indium-gallium-zinc oxide, IGZO).[19, 20] Electrochemical fibre transistors are typically assembled using shadow mask for metal evaporation to define the channel length ($L_c$) between electrodes.[20, 21] Semiconductors are coated over the electrodes of the electrochemical fibre transistor and ionic liquid is drop cast onto the semiconducting channel to enable gating of the device.[22] In contact with the ionic liquid, a metal wire is often used as the gate electrode.[15, 21] Maskless approaches have also been used to define the source and drain electrodes through the mechanical weaving of metal fibre.[15] Removing the lithography manufacturing step has a cost reduction advantage however, it is difficult to form a narrow $L_c$ < 0.5 mm which is needed for higher transistor performance and integration density.[15, 23] Field effect transistors (FETs) are typically assembled by using a metal wire to define a gate, [24-27] which is then coated with a dielectric material (e.g. aluminium oxide, parylene C) by evaporation.[24-26, 28] The semiconductor is then evaporated, coated or drop cast onto the dielectric.[24-29] Like electrochemical devices, metal evaporation through a shadow mask defines source and drain electrodes. The performance of the fibre transistors ranges from mobilities $\mu$ ~ 0.001 – 1.7 cm$^2$ V$^{-1}$ s$^{-1}$ with $I_{on}/I_{off}$ ~ 10$^3$ -10$^4$ for organic conducting polymers,[25-30] $\mu$ ~ 0.5 cm$^2$ V$^{-1}$ s$^{-1}$ and $I_{on}/I_{off}$ ~ 10$^4$ for single-walled carbon nanotubes (SWCNT)[18] and $\mu$ ~ 1.5 – 5.5 cm$^2$ V$^{-1}$ s$^{-1}$ and $I_{on}/I_{off}$ ~ 10$^4$-10$^7$ for IGZO.[19, 20, 31, 32] SWCNT transistors have excellent resilience to tensile strain >3% but it is difficult to isolate semiconducting SWCNT during production which can drive the cost >$1000 per gram.[33, 34] IGZO has demonstrated some of the best performances to date however, the $\mu$ decreases significantly to <0.1 cm$^2$ V$^{-1}$ s$^{-1}$ at <0.1% tensional strain making it challenging to implement in wearable products.[19] Organic conducting polymers such as P3HT do not suffer a $\mu$ drop with high strain (30-50%).[30] However, after several decades of research, they have reached a performance plateau $\mu$ ~ 10 cm$^2$ V$^{-1}$ s$^{-1}$, even in thin films.[35]

Transition metal dichalcogenides (TMD) flakes offer enormous potential and a new route forward in the development of fibre transistors offering a significant number of advantages such as flexibility,[11] durability,[5] minimal toxicity,[36] and transistors with flake networks having $\mu$ > 10 cm$^2$ V$^{-1}$ s$^{-1}$.[37, 38] Furthermore, TMD flakes can be made by liquid phase exfoliation (LPE) or electrochemical exfoliation (EE) and processed into the form of inks that can be readily applied on demand and provide benefits such as low-temperature processing (<120 °C), form factor benefits (e.g., easy to apply on rough >1 μm textile surfaces), lower production costs compared to growth methods[39], and scalability.[40] Despite high mobility in the basal plane of the LPE 2D flakes,[41] inter-flake junctions have typically limited the flake network $\mu$,[42] leading to low $\mu$ <0.3 cm$^2$ V$^{-1}$ s$^{-1}$.[43, 44] A new method of EE has recently emerged to produce large (aspect ratio, AR >100) semiconducting 2D flakes using quaternary ammonium salts (QAS).[38] The flakes can be used to create conformal flake networks that are limited by the quality of the 2D flake basal plane, enabling a route for network $\mu > 10$ cm$^2$ V$^{-1}$ s$^{-1}$.[37, 45] However, to our knowledge, fibre-based transistors with semiconducting flakes are yet to be achieved despite the tremendous potential. Furthermore, the biocompatibility of flakes made by QAS is largely unknown but extremely important if implemented in electronic fibre applications which could encounter human cells (e.g. electrodes, sensors, thermal interface components).

This work will demonstrate the effect of EE with MoS$_2$ and WSe$_2$ flakes on cytotoxicity with human skin cells to enable the use of 2D flakes in wearable applications. We will use EE flakes with polyester fibres to make conformal junctions and realise high $\mu$ WSe$_2$ and MoS$_2$ fibre transistors. Additionally, we will demonstrate a methodology to make fibre transistors with a short $L_c$ <200 μm without the need for lithography masks, to facilitate cost reduction, increase device density and enable a more straightforward manufacturing process towards high-performance fibre transistors.

**RESULTS:**

**Electrochemical exfoliation of TMD flakes**

We use EE to intercalate and expand bulk crystals of MoS$_2$ and WSe$_2$ using QAS in propylene carbonate (see Methods). The expanded crystals are ultrasonicated in a dispersion of poly(vinylpyrrolidone)/dimethylformamide (PVP/DMF) and subsequently centrifuged at 97$g$ to remove the unexpanded crystal. Next, centrifugation washing and solvent exchange were used to transfer the MoS$_2$ and WSe$_2$ flakes to isopropyl alcohol (IPA) to make our MoS$_2$ and WSe$_2$ ink, as described in the Methods section. IPA was chosen as it is a low boiling point solvent (~82.5°C), which will evaporate away quickly once deposited on the polyester fibre.[46]

In Figure 1a, we investigate the structure of the MoS$_2$ and WSe$_2$ flakes using X-ray diffraction (XRD) undertaken on drop-cast flake networks on silicon/silicon dioxide (Si/SiO$_2$) substrate. Both MoS$_2$ (green curve) and WSe$_2$ (brown curve) show diffraction peaks attributed to the 2H semiconducting phase of each material. [47-49] Two diffraction peaks located at 33° and 69° are also seen which we attribute to the silicon substrate.[50] Raman spectroscopy is used to determine the phase of the flakes in Figures 1b. Drop cast flake networks of $MoS_2$ (green curve) and $WSe_2$ (brown curve) showed the $A_{1g}$ and $E_{2g}$ Raman modes as expected[51, 52] and are consistent with previous reports of 2H semiconducting flakes. The J2 and J3 vibration modes attributed to the metallic 1T phase (located at 224 and 289 cm$^{-1}$ for $MoS_2$ and 218 and 236 cm$^{-1}$ for $WSe_2$) are absent.[53, 54]

We estimate the lateral flake size ($L$) and flake apparent thickness ($t$) using atomic force microscopy (AFM) statistics (See Methods). Figure 1c plots $L$ versus $t$ for individual $MoS_2$ and $WSe_2$ flakes, and no apparent correlation is observed. The average flake lateral size $<L>$ is 2.3±0.1 μm and 1.20±0.02 with an average apparent flake thickness $<t>$ of 9.7±0.2 nm and 17.4±0.5 nm for $MoS_2$ and $WSe_2$ respectively. Figure 1d is an AFM micrograph with an associated cross-section of a $WSe_2$ flake. Therefore the average flake aspect ratio ($<AR>$, $<L>/<t>$) is 251 and 98 for $MoS_2$ and $WSe_2$, respectively. These values are higher than the minimum AR required to make conformal flake-to-flake junctions (AR >40) which should also permit the flakes to wrap around fibres once deposited.[45]

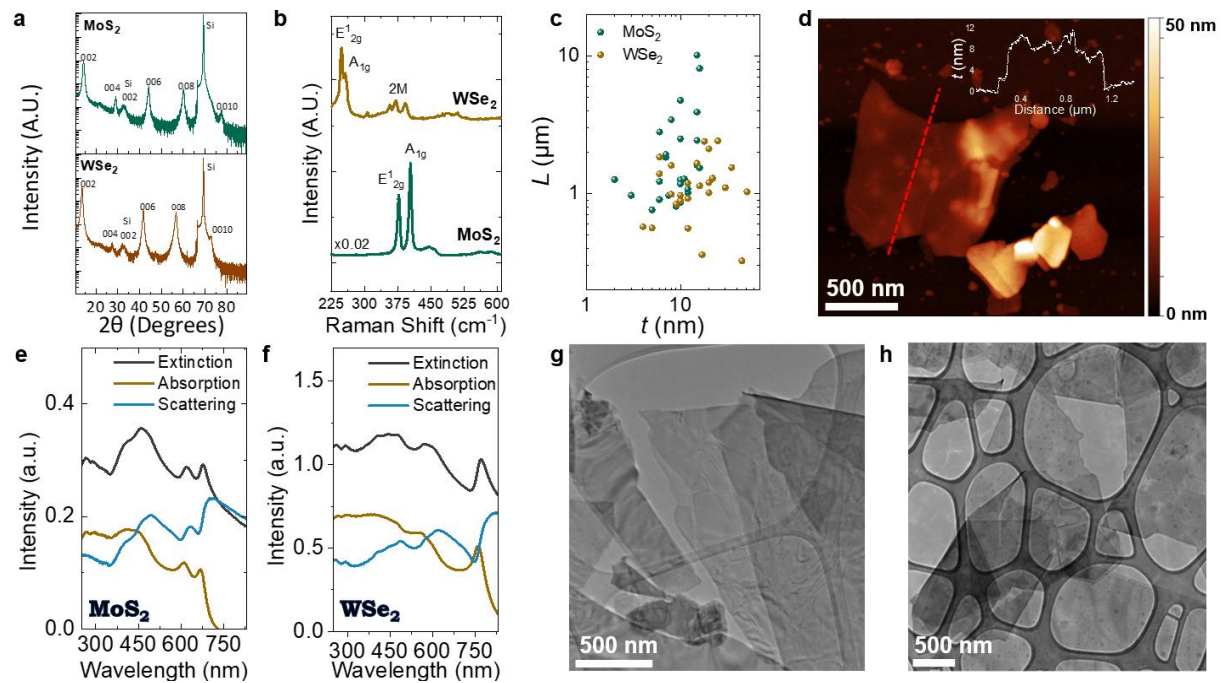

**Figure 1: Electrochemical Exfoliation of TMDs**. **a** XRD of the TMDs flakes demonstrating the semiconducing (2H phase) of the materials. **b** Raman spectroscopy chemical analysis of the TMDs following centrifugation into inks. **c** Atomic force microscopy statistics of the flake lateral size and thickness. **d** Atomic force microscopy micrograph of a representative $WSe_2$ flake. **e,f** UV-vis optical analysis of the TMD inks demonstrating the extinction, absorption, and scattering components as a function of wavelength for the $MoS_2$ and $WSe_2$ ink. **g,h** Transmission electron microscopy images of the $MoS_2$ and $WSe_2$ flakes showing $L > 1$ μm, flake fold and wrinkles.

Optical extinction, absorption, and scattering spectra of the $MoS_2$ and $WSe_2$ inks are presented in Figure 1e. Each spectrum shows the appropriate features with an A exciton at around 676 nm and 772 nm for $MoS_2$ and $WSe_2$, respectively, consistent with previous reports of LPE and EE flakes.[37, 55, 56] The scattering component of the ink is ~50% of the extinction spectra intensity and would suggest that the flakes have a large $L$.[55] Transmission electron microscopy (TEM) shown in Figure 1g and 1h also confirms the presence of large $L > 1$ μm flakes in the $MoS_2$ and $WSe_2$ inks.

**Electrochemically gated TMD fibre transistors**

Next, we test the performance of the semiconducting TMD inks by fabricating electrochemically gated fibre transistors. Most fibre transistor literature uses evaporation to deposit the semiconducting layer[25, 26, 28, 30], requiring the rotation of the fibre for complete coverage. Here, the TMD inks have a form factor benefit as the fibre does not need to be rotated during deposition, maximising the process' potential scalability. The $MoS_2$ and $WSe_2$ inks are drop cast (~140 μL, ~ 2 mg mL$^{-1}$) onto separate polyester fibres (Axel Suijker Textil, ⌀ ~ 200 μm diameter), composed of several hundred fibrils (⌀$_f$ ~ 10 μm) to assist capillary action coating along the fibre. There are ~ 400 fibrils per fibre ((⌀/⌀$_f$)$^2$). The $MoS_2$ and $WSe_2$ semiconducting fibres are created by annealing the coated fibres at 120 °C for 1 h on a hot plate in an $N_2$ glovebox to remove residual solvent and improve the adhesion of the $MoS_2$ and $WSe_2$ flake networks to polyester. In Figure 2a, scanning electron microscopy (SEM) imaging of the $MoS_2$ flake networks on the fibril surface reveals conformal inter-flake junctions between the flakes on the surface of the polyester fibrils, implying low junction resistance. Such conformal junctions are extremely important as they imply low junction resistance. It is worth noting that this is an important result as it implies that high aspect ratio semiconducting nanosheets prefer lie flat if it is possible and do not need special deposition conditions (e.g. Langmuir-Schaeffer deposition)[37], to enable them to align. In Figure 2b, cross-sectional SEM also reveals that the $MoS_2$ network of flakes has coated the circumference of the fibrils while the inside of the polyester fibrils remains uncoated, insulating and therefore bright in the SEM. Focused-ion beam SEM (FIB-SEM) imaging (see Methods) shown in Figure 2c determines a flake network thickness, $t_c$ = 242 ± 24 nm around the fibre circumference. The mechanical properties of the semiconducting fibres are tested in Figure 2d, which shows that a similar stress-strain curve is obtained for each fibre. An elastic modulus (slope of the stress-strain curve) of 0.57 ± 0.04, 0.31 ± 0.04 and 0.43 ± 0.13 GPa is found for the polyester (black curve), $MoS_2$/polyester (green curve) and $WSe_2$/polyester (brown curve) fibres respectively which is similar to previous reports for polyester fibres.[57] Therefore, the $MoS_2$ and $WSe_2$ flake networks are not significantly changing the mechanical properties of the polyester fibre. We assemble a fibre transistor by manually wrapping copper wire (⌀ = 100 μm) around the bundle of semiconducting fibres to define the source and drain electrodes. We also add a copper side-gate by wrapping a copper electrode ~ 1 cm from the channel, as shown in Figure 2e. The side gate is made of ~ 10 copper wires manually wound together to create an active diameter of ~ 1 mm. The gate is thicker than the source and drain electrodes, as previous work has shown that $\mu$ of ionic gated devices in 2D flake networks can be maximised when the gate volume is >10 times larger than the channel volume.[43] To complete our electrochemical transistor, shown in Figure 2e, we add a drop-cast ionic liquid (~50 μL) 1-ethyl-3-methylimidazolium bis(trifluoromethylsulfonyl)imide (EMIM TFSI) to allow gating of the semiconducting channel.[22] The electrochemical transistors are then characterised using a probe station in ambient air. In Figures 2f and 2g, we sweep a gate voltage ($V_g$) from -3V to 3V and apply a drain-source voltage, $V_{ds}$ = 1 V, to obtain the transfer characteristics for the $WSe_2$ and $MoS_2$ transistors, respectively. In Figure 2f, the $WSe_2$ curve demonstrates ambipolar behaviour with an off state at $V_g$ ~ 0 V while the $MoS_2$ transistor (Figure 2g) is n-type because the device turns off at negative $V_g$, confirming the electrical behaviour is as expected and consistent with previous reports for TMD flake networks on polyethylene terephthalate (PET).[37] We also observe that the gate leakage, $I_g$ attributed to the conductivity of the ionic liquid, is minimal (< 0.1 mA). The $\mu$ of the transistors is calculated from the equation $\mu = (L_c/W)(1/C_{device})(g_m/V_{ds})$, where $g_m = \partial I_d/\partial V_g$ is the transconductance, and $C_{device}$ is the device capacitance. We have 400 fibrils in the channel with source and drain electrodes separated by a channel length $L_c$ ~ 500 μm. For each fibril, their contribution to the channel width is the circumference of the fibril ($\pi\varnothing_f$), therefore $W = n\pi\varnothing_f$ where $n$ is the number of fibrils. Our previous work has estimated an average volumetric capacitance, $C_v \sim 1.4$ F cm$^{-3}$ for MoS$_2$ and WSe$_2$.[37] Using a flake network thickness $t_c = 242 \pm 24$ nm we can estimate $C_{device} = C_v \times t_c \sim 33.9$ μF cm$^{-2}$. Therefore, the average electron $\mu$ for MoS$_2$ and WSe$_2$ is calculated to be $\mu_{MoS2} \sim 0.9 \pm 0.1$ cm$^2$ V$^{-1}$ s$^{-1}$ and $\mu_{WSe2} \sim 0.8 \pm 0.3$ cm$^2$ V$^{-1}$ s$^{-1}$ with $I_{on}/I_{off} \sim 10^2$ and $\sim 10^4$ for MoS$_2$ and WSe$_2$ respectively. As an additional step we test the flexibility of the MoS$_2$ fibre transistor by bending it at a fixed bending radius between 2.5 - 5mm (see Methods) which varies the tensile strain, shown in the inset of figure 2g. We find $\mu$ decreases to 0.3 cm$^2$ V$^{-1}$ s$^{-1}$ at 4% tensile strain but remained operational.

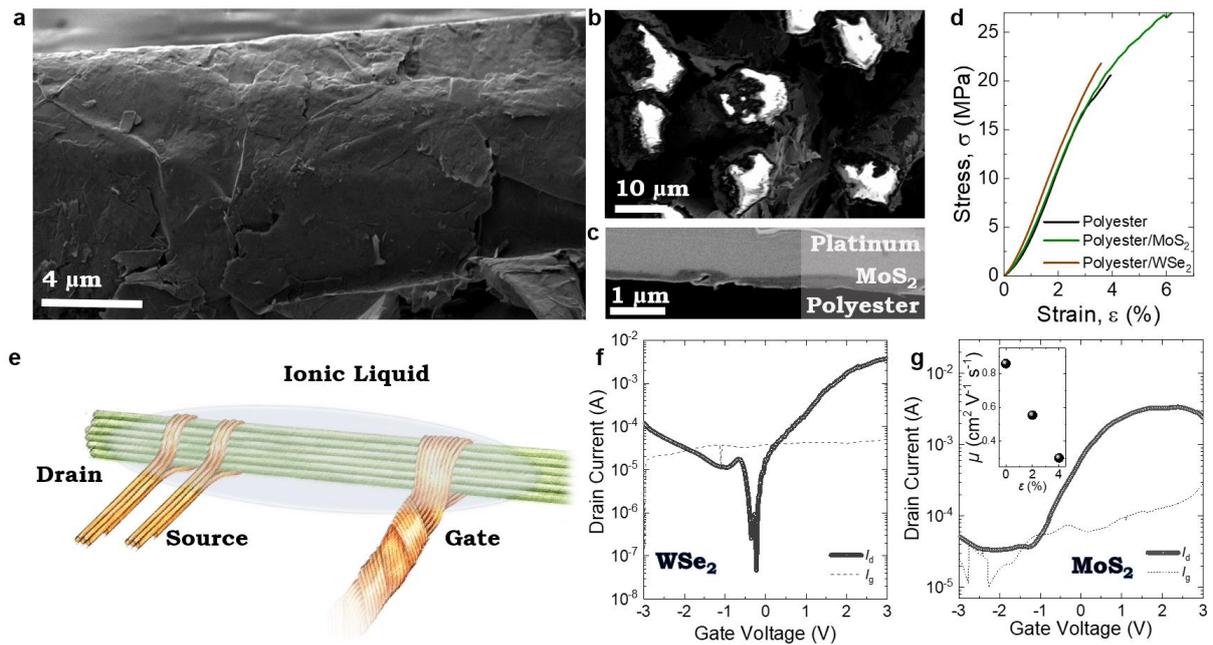

**Figure 2: Investigation of fibre morphology and electrical properties**. **a** SEM of a MoS$_2$ coated MoS$_2$ fibre demonstrating conformal junctions and morphologically uniform coating. **b** cross-sectional image of a MoS$_2$ coated polyester fibre showing an uncoated centre. **c** SEM-FIB of an MoS$_2$ coated fibril. **d** Mechanical testing of the TMD coated fibres. **e** Schematic of fibre transistor showing the MoS$_2$ coated fibres in green and the copper electrodes in brown. **f** Transfer characteristics of the WSe$_2$ and **g** MoS$_2$ coated fibres with the gate leakage shown as a dashed line. The $\mu$ as a function of tensile strain on the MoS$_2$ coated fibre is shown in the inset.

The $\mu$ of our TMD fibre transistors is greater than the majority of organic conducting polymer and SCNT fibre transistors with comparable $I_{on}/I_{off}$ [18, 25-30] and slightly less than IGZO fibres $\mu \sim 1.5 - 5.5$ cm$^2$ V$^{-1}$ s$^{-1}$ but with greater resilience to fibre strain (~4%), an essential requirement for fibre electronics.[19, 20, 31, 32] To our knowledge, electronic fibres with TMD's had not been achieved to date demonstrating an important first step. We attributed the reduction of the $\mu$ compared to planar devices on PET and Si/SiO$_2$ ($\mu$ > 10 cm$^2$ V$^{-1}$ s$^{-1}$) to the increased roughness of the fibrils and possible parasitic resistance in the channel created by the thick semiconducting layer $t_c$ = 242 nm [37, 38].

**Knotted Fibre TMD transistor**

To establish a route toward higher-performance fibre transistors ($\mu$ > 10 cm$^2$ V$^{-1}$ s$^{-1}$), patterning the source and drain electrodes needs a long-term strategy to decrease $L_c$ and $W$ < 500 μm for increased integration density, facilitate cost reduction and enable a straightforward manufacturing process without the need for lithography, which traditionally has been intended for use with fibres transistors. Kim et al. have used a twisted fibre assembly to overcome the difficulties of forming a narrow gap between source and drain electrodes. Gold fibre is coated with P3HT, and the coated fibres are twisted together to make $L_c$ as low as 120 nm achieving $\mu$ ~ 0.2 cm$^2$ V$^{-1}$ s$^{-1}$.[17] However, getting short channel $W$ < 4 cm in this configuration can be difficult. [17] Cross-bar geometry has also been used with a PEDOT:PSS coated polyamide fibre as the source, drain and semiconductor channel as one fibre and a second PEDOT:PSS coated fibre is positioned perpendicularly as the gate electrode.[12] The $L_c$ and $W$ are defined by the gate fibre diameter ~ 10 μm achieving $g_m$ ~ 10$^{-4}$ S however, the strategy is not viable for TMD materials where the semiconductor conductivity is too low ~ 10$^{-5}$ - 10$^0$ S/m to be used as a source-drain electrode.[12]

We demonstrate a new architecture to fabricate the fibre transistors, which use the fibre diameter (⌀ ~ 200 μm) to define $W$ and $L_c$. In Figure 3a, we use an overhand knot to wrap the semiconducting MoS$_2$ fibre around a copper wire (⌀ = 100 μm) to define the drain. Next, a second overhand knot is used to position the source. A force (F) is applied by hand in the direction of the arrows to tighten the MoS$_2$ fibre around the source and drain electrodes. As a result, the source-drain electrodes do not short-circuit since the MoS$_2$ fibre is between them. Assuming the the current will only flow from the copper source to drain fibre we define $L_c$ = 2×⌀ = 400 μm, twice the fibre diameter and $W = n\pi⌀_f$ = 12.57 mm. Next, we add a copper side gate by wrapping a copper wire ~ 1 cm from the channel, with a diameter of ~ 1 mm. EMIM TFSI is drop cast (~ 50 μL) over the channel to gate the device.

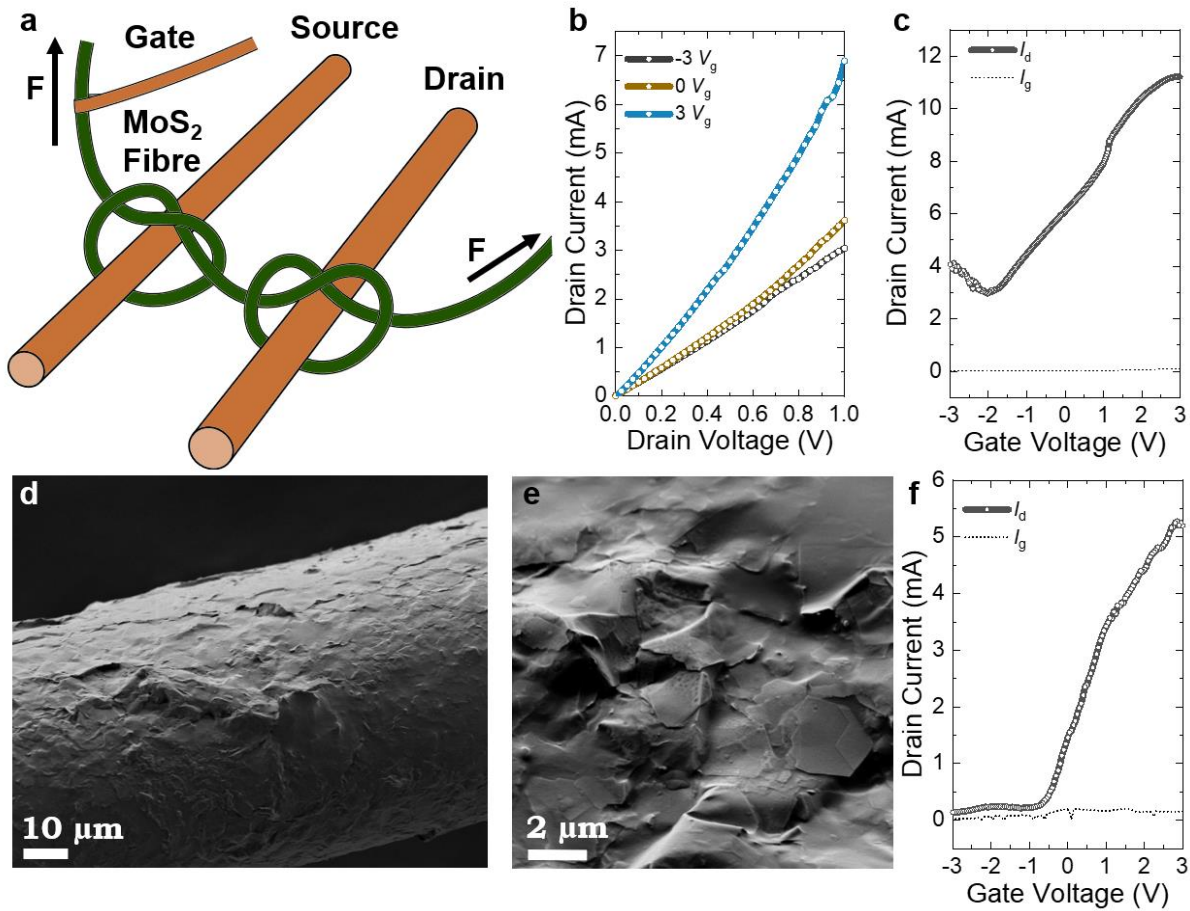

**Figure 3: Investigation of knot architecture fibre transistors and their electrical properties. a** Schematic of fibre knot fibre transistor showing the $MoS_2$ coated fibres in green and the copper electrodes in brown. A force (F) is applied by hand in the direction of the arrows to tighten the $MoS_2$ fibre around the source and drain electrodes. **b** Output characteristic of a $MoS_2$ knot fibre transistor. **c** Transfer characteristics of the $MoS_2$ knot fibre transistor with the gate leakage shown as a dashed line. **d** SEM of a human hair with $MoS_2$ flakes coated onto the circumference **e** Magnified area of the $MoS_2$ coating on the human hair showing conformal flake-to-flake junctions **f** Transfer characteristic of the human hair transistor using the knot architecture. Gate leakage is shown as the dashed line.

In Figure 3b, we examine the output characteristics of the $MoS_2$ knot-based fibre transistor at gate voltages of 3 V, 0 V, and -3 V. These curves are consistent with our previous $MoS_2$ fibres showing n-type behaviour and turn off at negative $V_g$ (black curve), shown as a drop in $I_d$ (<3 mA at 1 $V_{ds}$). In addition, we find our average transistor $\mu \sim 0.7 \pm 0.3$ $cm^2$ $V^{-1}$ $s^{-1}$ from the slope in the linear region of the transfer characteristic, shown in Figure 3c, which is comparable to our previous MoS$_2$ fibre devices but with a reduction of $I_{on}/I_{off} \sim 10$ potentially due to charge screening. As an additional step towards decreasing the channel length, we fabricate an additional MoS$_2$ transistor with the knot architecture using the same manufacturing protocol but replacing the polyester for a human hair with $\varnothing = 70$ µm. Like the muti-fibril polyester fibre, we observe a homogeneous coating of MoS$_2$ flakes around the circumference of the fibre (Figure 3d) and conformal flake-to-flake junctions (Figure 3d) from SEM images. In this case, $L = 2 \times \varnothing = 140$ µm and $W = n\pi\varnothing = 220$ µm since $n = 1$. Therefore, we find $\mu \sim 15 \pm 6$ cm$^2$ V$^{-1}$ s$^{-1}$ and $I_{on}/I_{off} \sim 26 \pm 4$ from the transfer characteristic shown in Figure 3f. We find a sizeable improvement in $\mu$ likely due to the reduced transistor dimensions of $L$ and $W$. Despite the reduction in $I_{on}/I_{off}$ the architecture represents a potential route to decrease $L_c$ and increase $\mu$ by shrinking the transistor geometry and reducing the number of flake-to-flake junctions, the limiting factor in flake networks. In this case, resolution is limited by our manual (by hand) assembly of the structure, however it might be possible to decrease $L_c$ further by using thinner fibre diameters such as electrospun fibres[58] using machine-based assembly with this architecture. A smaller fibre diameter might also help reduce parasitic channel resistance and charge screening, improving device performance. Although small diameter fibres ($\varnothing < 10$ µm) are out of the scope of this work, the fabrication of high mobility $\mu > 10$ cm$^2$ V$^{-1}$ s$^{-1}$ fibre transistors with TMD flakes represents an important first step for the field.

**Biocompatibility of TMD Inks**

The cytotoxic effect of TMD flakes on human cell lines after EE with QAS is largely unknown and can vary depending on concentration, shape, exposure time, functionalisation and cell line being tested.[59, 60] LPE flakes of MoS$_2$ have demonstrated low cytotoxicity with a keratinocyte cell line for at least 48 h.[61] Conversely, the viability (using a colorimetric assay) of human lung carcinoma epithelial cells drops up to 60% after 24 h of exposure to MoS$_2$ flakes produced by EE using Li-intercalation.[62] Therefore, as a further investigation, we investigate the biocompatibility of our MoS$_2$ and WSe$_2$ flakes to enable real-world applications of the electronic fibres that may be exposed near human skin. We choose keratinocytes to model the skin sub-surface. Before testing, the WSe$_2$ and MoS$_2$ inks were solvent exchanged into deionised water (see Methods). The keratinoxyte cell bodies and nuclei are are stained with phalloidin and DAPI and fluorescently imaged (see Methods). In Figure 4a and 4b, the DNA content and cell metabolic activity were monitored after 80 µg mL$^{-1}$ MoS$_2$ and 60 µg mL$^{-1}$ WSe$_2$ ink addition for 24 and 72 h, respectively. No significant change in the DNA content and metabolic activity per DNA content was observed, indicating the biocompatibility of the MoS$_2$ and WSe$_2$ flakes for at least 72 h. WSe$_2$ also showed a potentially beneficial effect on the viability of the keratinocytes, with an increase in DNA content and metabolic activity per cell for the WSe$_2$-treated cells, possibly due to the presence of selenium which is an essential element for keratinocyte function.[63] Fluorescence imaging of the cells (see Methods) after 3 days in culture with the MoS$_2$ (Figure 4c) and WSe$_2$ (Figure 4d) flakes at concentrations between 40 – 80 µg mL$^{-1}$ yielded cells with robust proliferation and healthy cellular morphology confirmed by the high nuclei count (Figure 4e) and area fraction (i.e. area covered by the cells) (Figure 4f). Finally, to simulate the longer-term interaction of cells with a TMD, keratinocytes were cultured with MoS$_2$ for 14 days. Fluorescent imaging demonstrated robust proliferation of the cells, with healthy cellular morphology (Figure 1g) that was similar to the media-only control (Figure 1h).

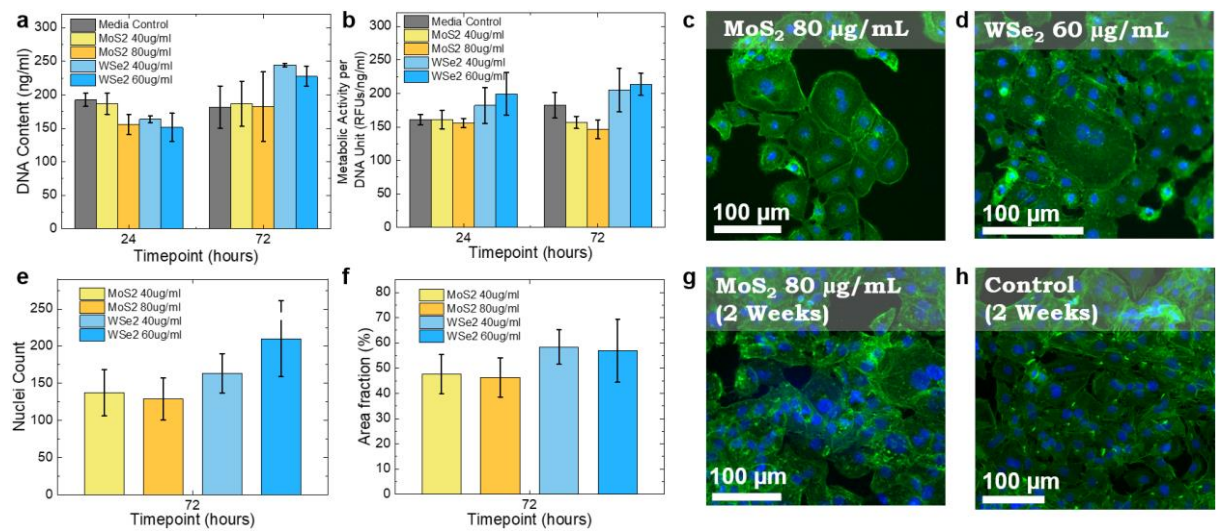

**Figure 4: Biocompatibility testing of TMDs**. **a** Quantification of the DNA content as a function of time. **b** Quantification of the metabolic activity per DNA unit as a function of time. **c,d** Phalloidin (green) and DAPI (blue) stained HaCaT keratinocytes after 72 h growth in suspension with 80 µg mL$^{-1}$ MoS$_2$ and 60 µg mL$^{-1}$ WSe$_2$. **e,f** Nuclei count and cell coverage found from microscopy after 72 h. **g**

Fluorescence image of keratinocytes grown in suspension with $MoS_2$ after 14 days of growth. **h** Fluorescence image of keratinocytes with no addition of TMD's after 14 days of growth.

**Conclusions:**

We find that high aspect ratio (AR >100) TMD flakes will self-align on polyester fibre after drop casting and make conformal flake-to-flake junctions that enable flexible high-mobility electronic networks $\mu \sim$ 15 cm$^2$ V$^{-1}$ s$^{-1}$, which has previously not been possible with TMD's in a fibre structure. $MoS_2$ fibre transistors were also demonstrated to be operational at a high tensile strain ~4%, a key advantage of 2D materials and an essential requirement for wearable devices. We propose a novel 'knot' architecture, a lithography-free and textile-compatible method of manufacture to scale down future fibre transistors to the length scale of twice the fibre diameter and demonstrate its operation with both multifibril polyester fibres and a human hair. As a final demonstration, electrochemical exfoliation of TMDs with QAS was biocompatible with human skin cells. Keratinocytes cultured with $MoS_2$ and $WSe_2$ showed high viability, proliferation and healthy cellular morphology over three days in culture and 14 days in culture for $MoS_2$. Therefore the TMDs are suitable candidates for electronic fibres that may come into contact with human skin broadening their impact in enhancing the integration of electronics with human daily life.

**Methods:**

**Electrochemical exfoliation of 2D crystals:** To intercalate $WSe_2$ and $MoS_2$ (HQ graphene) crystals, a two-electrode electrochemical cell is utilised. The cathode is a thin piece of crystal measuring 0.1 × 1 × 1 mm, while a platinum foil from Alfa Aesar serves as the anode. The electrodes are held in place with copper crocodile clips. The electrolyte consists of 5 mg mL$^{-1}$ tetrapropylammonium (TPA) bromide from Sigma-Aldrich added to 50 mL of propylene carbonate. A voltage of 8V is applied between the electrodes for 30 minutes to intercalate the 2D crystal with TPA+ cations. The expansion of the 2D crystal to more than twice its original volume in each case confirms the successful intercalation of the crystal.

**Ink formulation with 2D crystals**: The expanded 2D crystal undergoes bath sonication (Fisherbrand 112xx series) using 1 mg mL$^{-1}$ PVP with a molecular weight of approximately 40,000 dispersed in DMF for 5 min. Afterwards, the mixture is centrifuged using a Hettich Mikro 220, with a radius of 87 mm, at 500 rpm (24g) for 20 minutes to remove unexfoliated crystals. The dispersion is then size-selected by centrifuging the supernatant (top 90%) at 1000 rpm (97g) for 1 h and collecting the sediment. To eliminate the PVP, the 97g sediment was diluted with 2 mL of DMF and centrifuged at 10,000 rpm (9744g). This process was repeated twice, and the sediment was collected each time. A third washing step was implemented to remove residual DMF, in which the sediment was diluted in IPA (0.5 mL) and centrifuged at 10k rpm (9744g), and the sediment was collected. The sediment was then redispersed in IPA (~0.5 mL) with a concentration of approximately 2.5 mg mL$^{-1}$ for each crystal to create the $MoS_2$ and $WSe_2$ inks, respectively.

**X-ray diffraction**: XRD spectra were captured using a PanAlytical X'Pert Pro diffractometer with a Cu tube emitting Kα radiation (1.5406 Å). The spectra were acquired in the 2θ range from 10° to 90° on the flake networks prepared on Si/SiO$_2$ (100) orientated wafers (~2000 nm oxide thickness).

**Transmission Electron Microscopy**: TEM lacey carbon grids were prepared for imaging by drop casting the $MoS_2$ and $WSe_2$ inks and leaving to dry overnight. A JEOL 2100 TEM was used for imaging the grids with a 200 kV accelerating voltage and 105 μA beam current.

**Scanning Electron Microscopy**: SEM imaging was conducted using a Carl Zeiss Ultra SEM. A secondary electron detector was used to obtain the images at a 1 and 3 kV accelerating voltage and 30 μm aperture.

**FIB-SEM Cross Section Imaging**: FIB-SEM microscopy is carried out using a dual-beam Carl Zeiss Auriga focused ion beam system. A platinum pad was deposited by dissociation of a organometallic gas by the electron beam. Network cross sections were milled with a 30 kV:120 pA gallium ion beam. All images were captured using the inlens detector with an accelerating voltage of 2 kV and a 30 μm aperture at a working distance of 5 mm. The flake network thickness is averaged from 18 measurements at different points along the coated fibre.

**Electrical Measurements**: We employ EMIM TFSI (Sigma Aldrich), an ionic liquid, to control ion injection into our semiconducting channel. The ionic liquid was heated at 70 °C under a vacuum (~$1.6 \times 10^{-4}$ mbar) overnight on Si/SiO$_2$ to remove any water that may have been absorbed. The ionic liquid EMIM was then drop cast across the fibre, covering the gate, source, and drain electrodes. The devices were connected to a Keithley 2612A dual-channel source measurement unit employing gold-coated probes in ambient air to conduct electrical characterisation. A scan rate of 50 mV/s, $V_{ds}$ = 1 V and a gate voltage window of -3 V to 3 V were used to assess the transfer characteristics.

**Atomic Force Microscopy:** We conducted AFM to measure the flakes' thickness and lateral dimensions using a Bruker Multimode 8 microscope. Before imaging we dilute (1:100) the MoS$_2$ and WSe$_2$ inks in IPA and drop-cast onto Si/SiO. The samples were then annealed for 15 min at 120°C in a Ar glovebox (UNIlab Pro, Mbraun) to remove any leftover solvent. The samples were scanned using the OLTESPA R3 cantilevers in ScanAsyst mode, 25 flakes were chosen for statistical analysis.

**Raman Spectroscopy:** Inks of MoS$_2$ and WSe$_2$ are drop cast onto Si/SiO$_2$ substrate and annealed at 120 °C. A WITec Raman spectrometer at 532 nm with a 20× objective is used to acquire spectra. An incident power of ∼1 mW was used to minimise possible thermal damage.

**Optical absorption spectroscopy**: A 10 mm optical length quartz cuvette and a Cary 1050 spectrometer were used to produce the spectra. The absorption spectra were collected in an integrating sphere. The collected extinction and absorption spectra of nanosheet dispersions were subtracted by their corresponding IPA spectra to yield flake-only spectra. The scattering spectra were obtained by using extinction spectra and subtracting absorption spectra.

**Biological testing:** The WSe$_2$ and MoS$_2$ inks were centrifuged at 10k rpm (9744g), and the IPA supernatant was decanted. The sediment was then redispersed in deionised water (~ 0.5 mL) and centrifuged at 10k rpm (9744g) to remove residual IPA. The IPA/water supernatant was decanted, and the sediment was redispersed in deionised water (~ 0.5 mL) at 2.5 mg mL$^{-1}$ concentration. HaCaT keratinocytes were cultured in suspension with 40 µg mL$^{-1}$, 60 µg mL$^{-1}$ or 80 µg mL$^{-1}$ of the respective TMDC, diluted in growth media (low glucose Dulbecco's Modified Eagle Medium, 1% penicillin/streptomycin, 1% l-glutamine and 10% foetal bovine serum (FBS)). The metabolic activity was assessed by the Alamar Blue assay (Invitrogen, UK), and the DNA content was assessed by the Picogreen assay (Invitrogen, UK) on days 1 and 3. Cells were fixed using 10% formalin in PBS on days 3 and 14. Cells were subsequently stained with phalloidin to label the cell membrane, DAPI to label the nucleus, and fluorescently imaged on a Zeiss AxioObserver microscope to investigate the cell morphologies. Analysis of cell counts and area fraction was carried out using FIJI.[64]

**Stress-strain and bending tests:** Tensile measurements were carried out with a Zwick Z0.5 ProLine Universal Testing Machine (100 N Load Cell) at a strain rate of 1 % per second. All samples had a gauge length, $L_0 = 20$ mm and were placed under a 500 kPa pre-load to ensure fibres were taught at the start of the measurement. For the $MoS_2$ FET bending tests, the device (⌀ ~ 200 μm) was mounted on PET (thickness = 140 μm) with kapton tape and bent using a fixed bending radius of 2.5 and 5 mm, respectively, strained parallel to the channel. Tensile strain ($\varepsilon$) was equal to $y/2r$, where $r$ is the bending radius and $y$ is the sample thickness.

**Acknowledgements:**

We acknowledge the European Commission (Graphene Flagship Core 2 and Core 3 grant agreement No. 785219 and 881603, respectively) and the European Research Council (FUTURE-PRINT). We have also received support from the Science Foundation Ireland (SFI) funded centre AMBER (SFI/12/RC/2278_P2) and availed of the facilities of the SFI-funded AML and ARL labs. C.I. acknowledges financial support of the Ministry of Education of Turkey and AMBER (Trinity College Dublin). T.C. acknowledge funding from a Marie Skłodowska-Curie Individual Fellowship "MOVE" (grant number 101030735, project number 211395, and award number 16883).

**Data Availability:**

The authors declare that the data supporting the findings of this study are available within the paper and its supplementary information files. Data is also available from the corresponding author upon reasonable request.